\documentclass{article}

\usepackage{arxiv}

\usepackage[utf8]{inputenc} 
\usepackage[T1]{fontenc}    
\usepackage{hyperref}       
\usepackage{url}            
\usepackage{booktabs}       
\usepackage{amsfonts}       
\usepackage{nicefrac}       
\usepackage{microtype}      
\usepackage{lipsum}		
\usepackage{graphicx}
\usepackage{natbib}
\usepackage{doi}
\usepackage{microtype}
\usepackage{graphicx}
\usepackage{subfigure}
\usepackage{booktabs}
\usepackage{algorithm}
\usepackage{algpseudocode}
\usepackage{natbib}

\usepackage{hyperref}
\usepackage{graphicx}
\usepackage{colortbl}
\definecolor{DarkGray}{gray}{0.8}
\usepackage{fancyhdr}

\title{Enhancing Situational Awareness in Surveillance: Leveraging Data Visualization Techniques for Machine Learning-based Video Analytics Outcomes}

\author{ \href{https://orcid.org/0000-0002-3936-4552}{\includegraphics[scale=0.06]{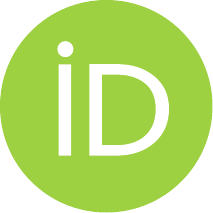}\hspace{1mm}Babak Rahimi~Ardabili}\\
	Department of Public Policy\\
	University of North Carolina at Charlotte\\
	Charlotte, NC 28223 \\
	\texttt{brahimia@charlotte.edu} \\
	 \AND
	 Shanle Yao \\
	 Department of Electrical and Computer Engineering \\
      University of North Carolina at Charlotte\\  
	 Charlotte, NC 28223 \\
	 \texttt{syao@charlotte.edu} \\
	\And
	 Armin Danesh~Pazho \\
	 Department of Electrical and Computer Engineering \\
  University of North Carolina at Charlotte\\
	 Charlotte, NC 28223 \\
	 \texttt{adaneshp@charlotte.edu} \\
	\And
	 Lauren Bourque \\
	 Department of Electrical and Computer Engineering \\
  University of North Carolina at Charlotte\\
	 Charlotte, NC 28223 \\
	 \texttt{lbourqu1@charlotte.edu} \\
     \And
	 Hamed Tabkhi \\
	 Department of Electrical and Computer Engineering \\
  University of North Carolina at Charlotte\\
	 Charlotte, NC 28223 \\
	 \texttt{htabkhiv@charlotte.edu} \\
}

\date{}

\renewcommand{\headeright}{}
\renewcommand{\undertitle}{}
\renewcommand{\shorttitle}{Enhancing Situational Awareness in Surveillance}

\hypersetup{
pdftitle={Enhancing Situational Awareness in Surveillance},
pdfsubject={q-bio.NC, q-bio.QM},
pdfauthor={Babak Rahimi~Ardabili},
pdfkeywords={Computer Vision, Smart Video Surveillance, Data Analysis, Data Visualization},
}

\begin{document}
\maketitle

\begin{abstract}
	The pervasive deployment of surveillance cameras produces a massive volume of data, requiring nuanced interpretation. This study thoroughly examines data representation and visualization techniques tailored for AI surveillance data within current infrastructures. It delves into essential data metrics, methods for situational awareness, and various visualization techniques, highlighting their potential to enhance safety and guide urban development.
This study is built upon real-world research conducted in a community college environment, utilizing eight cameras over eight days. This study presents tools like the Occupancy Indicator, Statistical Anomaly Detection, Bird's Eye View, and Heatmaps to elucidate pedestrian behaviors, surveillance, and public safety. Given the intricate data from smart video surveillance, such as bounding boxes and segmented images, we aim to convert these computer vision results into intuitive visualizations and actionable insights for stakeholders, including law enforcement, urban planners, and social scientists. The results emphasize the crucial impact of visualizing AI surveillance data on emergency handling, public health protocols, crowd control, resource distribution, predictive modeling, city planning, and informed decision-making.
\end{abstract}


\section{Introduction}\label{Intro}

The rapid evolution of technology has profoundly reshaped the realm of security and surveillance. The extensive installation of Closed-Circuit Television (CCTV) cameras has become a hallmark of modern urban settings, serving as ever-watchful guardians monitoring public areas, vital infrastructures, and private properties. As these cameras multiply, cities and institutions grapple with an overwhelming influx of video data. In 2023, the video surveillance sector was worth USD 53.7 billion and is predicted to exhibit a Compound Annual Growth Rate (CAGR) of 9.2\%, surpassing the broader IT sector's average CAGR of 5\%, with an anticipated valuation of USD 83.3 billion by 2028\footnote{ https://www.marketsandmarkets.com/}. Yet, this data deluge presents unique challenges.

A primary concern is the effective oversight and interpretation of these video outputs \cite{dilshad2020applications}. Legacy video surveillance frameworks, dependent on human operators for manual video review, are becoming less feasible for two main reasons. Firstly, the vastness of video content renders comprehensive and effective simultaneous monitoring a daunting task \cite{tang2021visualization}, especially when pinpointing specific incidents amid hours of footage. Secondly, many global law enforcement bodies face personnel deficits \cite{de2023insights}, resulting in an inadequate workforce to oversee the expanding camera networks. This scenario underscores the imperative to transition to intelligent surveillance systems capable of independent video analysis and delivering meaningful insights.

\begin{figure}
        \centering
               \includegraphics[width=0.8\linewidth, trim= 0 110 0 0,clip]{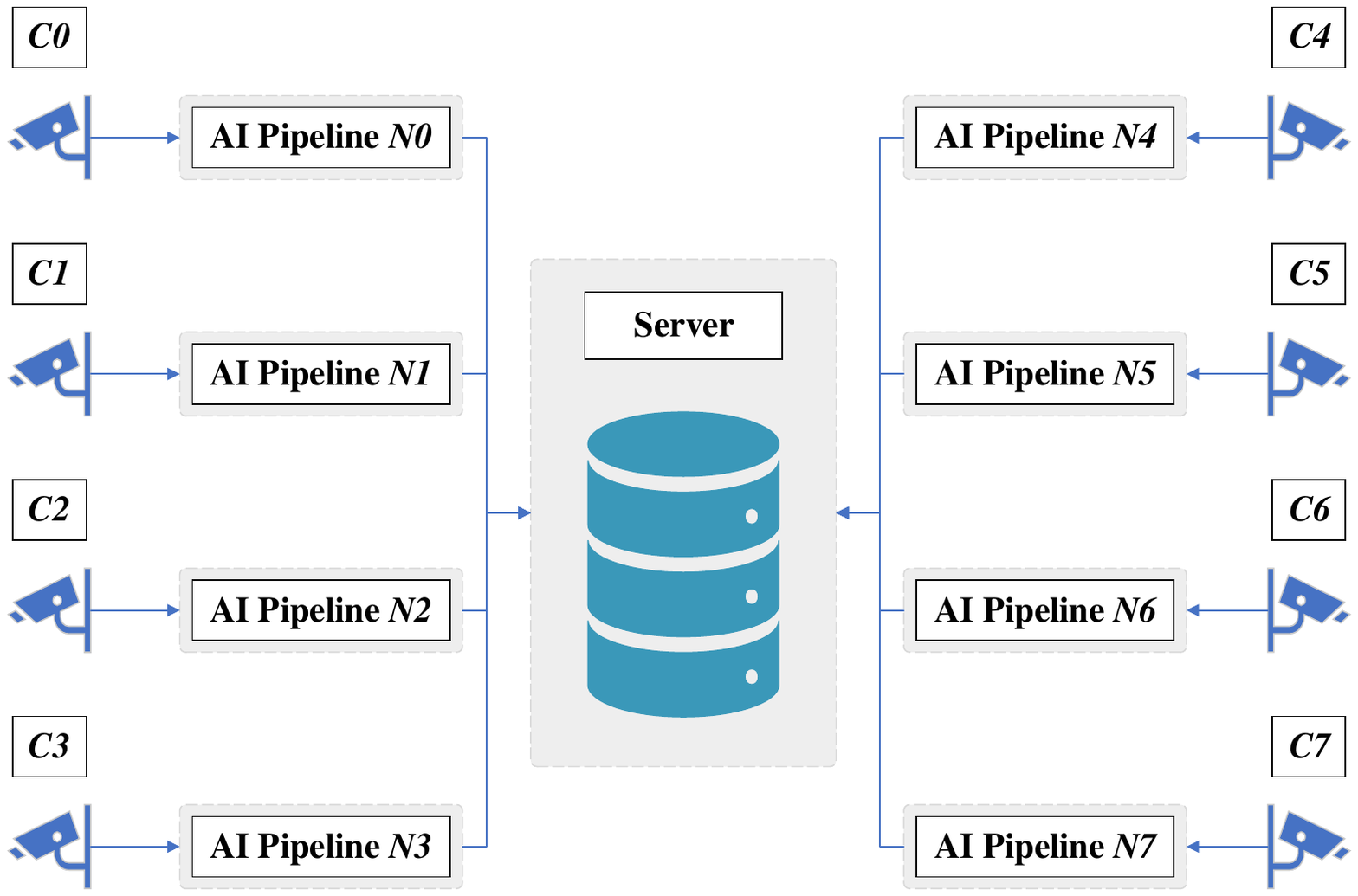}
                    \caption{Conceptual Graph illustrating the interconnected structure of multiple AI pipelines across various nodes and the database.}
                    
                \label{fig:Conceptual}
\end{figure} 

As cameras become more ubiquitous, there's growing public apprehension about the potential misuse of these systems, leading to unwarranted surveillance and potential infringement of individual privacy rights \cite{ardabili2022understanding}. These ethical and privacy issues regarding facial recognition technologies motivated many researchers to use more abstract approaches in video analysis \cite{pazho2023ancilia, ardabili2023understanding, noghre2023understanding}.  

Using this approach in video analytics generated a huge volume of meaningless data for the stakeholders \cite{myagmar2023survey}. This data needs to be analyzed and transformed into meaningful data. Therefore, big data analysis has become pivotal in this area. Time and space-related data provide a chronological sequence of observations, which can offer valuable insights into patterns and trends in the context of surveillance \cite{xu2016big}. Furthermore, visualization techniques can play a crucial role in synthesizing and presenting this data in a manner that's intuitive and actionable \cite{cauti2023intelligent}. 

This paper sheds light on translating computer vision outputs, such as bounding boxes, crop features, and re-identifications, into human-understandable insights for various stakeholders and decision-makers. These stakeholders encompass law enforcement and campus security personnel seeking to maintain safety and security in monitored areas as parking lots or shopping malls, social scientists exploring human behavior variations, architects and urban designers interested in comprehending human mobility, transportation planners, and even marketing analysts delving into customer behavior within these spaces.
To this end, we leverage our real-world testbed at a community college, collecting real-time data over eight days, comprising four holidays and four weekdays, across eight CCTV cameras. Leveraging the existing infrastructure of the educational institution, we harnessed eight cameras, including two outdoors and six indoors, to gather essential data. Within this framework, our machine learning-based video analytics relied on privacy-preserving pose-based algorithms, a departure from pixel-based models \cite{pazho2023ancilia}. The conceptual structure of this system is depicted in Figure \ref{fig:Conceptual}.

Our approach employs data analysis techniques to generate statistical outputs that assist human observers in comprehending historical data and trends. First, we used statistical analytics to provide stats such as headcounts, the average number of people detected hourly, and the total number of people over time to provide general overview of the raw AI data. Furthermore, we employ visualization techniques on top of these statistical outputs to enhance situational awareness. For instance, by extracting and analyzing data such as bounding boxes and human trajectories, we applied a transformation matrix considering the camera angle to get the bird's eye view at specific times. Appending these bird's eye view data, we developed a heat map representation of spatial movement over time to identify areas of usage and underutilization. We also utilized the historical data of each camera to generate occupancy indicators that represent a meaningful visualization of current people at each location regarding the historical data. Also, we present the statistical anomalies by extracting the normal distribution of the data trend per location.  In summary, the main contributions of this paper are:
\vspace{-5pt}
\begin{itemize}
\item We introduce an added layer of statistical analysis to computer vision outputs, such as bounding boxes, segmented images, and re-identifications. This layer helps extract significant insights from the foundational computer vision data.
\item We employ visualization techniques to render the statistical findings in a manner that's easily comprehensible to human observers and stakeholders. This approach amplifies the data's accessibility and usability, providing a lucid grasp of the intricate details procured from the computer vision system, thereby boosting situational awareness.
\item We delve into the valuable insights extracted from data visualization, emphasizing their potential role in augmenting safety measures.
\end{itemize}

\vskip -0.1in
Following this introduction, the paper is organized into distinct sections to provide a comprehensive insight into the topic. In Section \ref{Lit}, we offer an exhaustive survey of existing research using data visualization techniques using machine learning-based methodologies outputs. After this, section \ref{Database} delves into the explanation of data collection, the database's architecture, and a closer look into the features. Section \ref{Data}  provides the results in two parts: \textit{Descriptive Data} and \textit{Situational Awareness}. Finally, Section \ref{Discuss} synthesizes the findings from the preceding sections, debating their potential implications and charting pathways for future research in this domain.

\section{Literature Review}\label{Lit}
Machine learning and artificial intelligence advancement in various domains have paved the way for complex data analysis and visualization. The criticality of analyzing the outputs of these machine learning models is evident in numerous applications, ranging from urban planning to safety considerations. However, the effectiveness of these models heavily relies on the presentation and interpretation of the analyzed data.

The advent of machine learning techniques, especially when combined with IoT devices, has revolutionized urban planning and smart cities. Mahdavinejad et al. underline the importance of such methods in forecasting urban congestion and paving the way for solutions \cite{mahdavinejad2018machine}. Complementing this, Zanella et al. argue for the accessibility of urban IoT data to authorities and citizens, fostering civic participation and swift responsiveness to urban challenges \cite{zanella2014internet}. Taking this a step further, Rathore et al. delve deep into the capacities of IoT systems for urban development, focusing on vehicular traffic datasets \cite{rathore2016urban}. However, the true potential of these systems is realized when combined with big data analytics, as highlighted by Al Nuaimi et al., who stress iterative improvements based on analytical feedback \cite{al2015applications}.

Transitioning from holistic urban planning, it's imperative to focus on specific aspects, like traffic and pedestrian safety, which are paramount for the efficient functioning of a city.

Safety remains at the forefront of urban considerations, and several studies have dedicated efforts to analyze traffic patterns and potential hazards. Mouchili et al. venture into vehicle traffic data, shedding light on anomalies and parking occupancy \cite{mouchili2018smart}. Abberley et al. expand this scope by using machine learning to study traffic accidents, revealing patterns similar to human traffic analytics \cite{abberley2017modelling}. Similarly, Bharadwaj et al. apply image processing techniques on highway images, concentrating on vehicle placements with a unique approach to data representation \cite{bharadwaj2016traffic}. The realm of pedestrian safety isn't untouched either, with Chen et al. employing machine learning on Street View images to detect pedestrians, albeit without delving into intricate visualizations \cite{chen2020estimating}. However, safety isn't confined to traffic. He et al. highlight the use of machine learning in detecting crime-facilitating factors in neighborhoods, emphasizing the dire need for robust visual representation for impactful insights \cite{he2017built}.

While these studies provide critical insights into urban planning and safety, how the data is presented plays a crucial role. The following studies underline the importance of data visualization and interpretation.
\begin{figure}
        \centering
               \includegraphics[width=0.6\linewidth, trim= 0 0 0 0,clip]{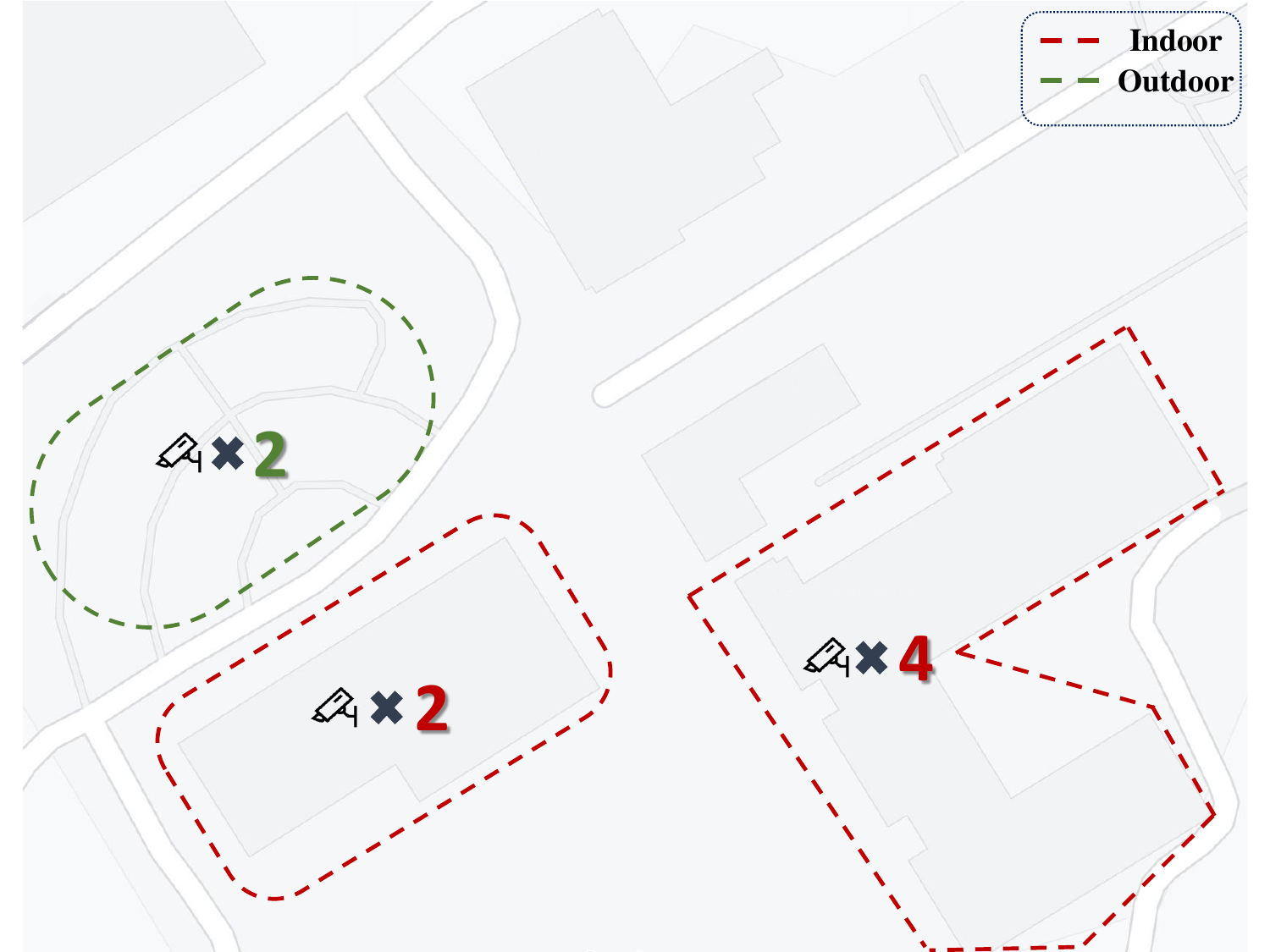}
               \vskip -0.05in
                    \caption{Camera locations in the study area. The red-dotted areas are two buildings, and the green-dotted area is the parking lot. The number of cameras in each location is presented per location. }
                    
                \label{fig:CL}
\end{figure}
Visualization techniques serve as the bridge between complex data and user comprehension. Soltani and Sardari vouch for the union of advanced technology and data processing algorithms to tackle challenges in computer vision systems, especially in defect detection \cite{soltani2022defect}. Similarly, Papadakis et al. introduce a computer vision system to detect fish species but underscore the need for advanced analytics for a holistic understanding \cite{papadakis2012computer}. Recognizing the importance of interpretability, Vellido emphasizes the necessity of user-friendly visualizations, especially in critical domains like medicine \cite{vellido2020importance}. Gonzalez et al. further this notion by presenting a biomedical image processing system that provides real-time data analysis and visualization \cite{gonzalez2019biomedical}. Leaping urban forestry, Cai et al. demonstrate the use of deep learning to quantify urban tree cover, offering visualizations that are both detailed and insightful \cite{cai2018treepedia}. Lastly, Gebru et al. masterfully employ deep learning on Street View images to deduce demographic patterns, setting a benchmark for the effective use of visualizations \cite{gebru2017using}.

\begin{figure*}
        \centering
               \includegraphics[width=1\linewidth, trim= 0 353 0 0,clip]{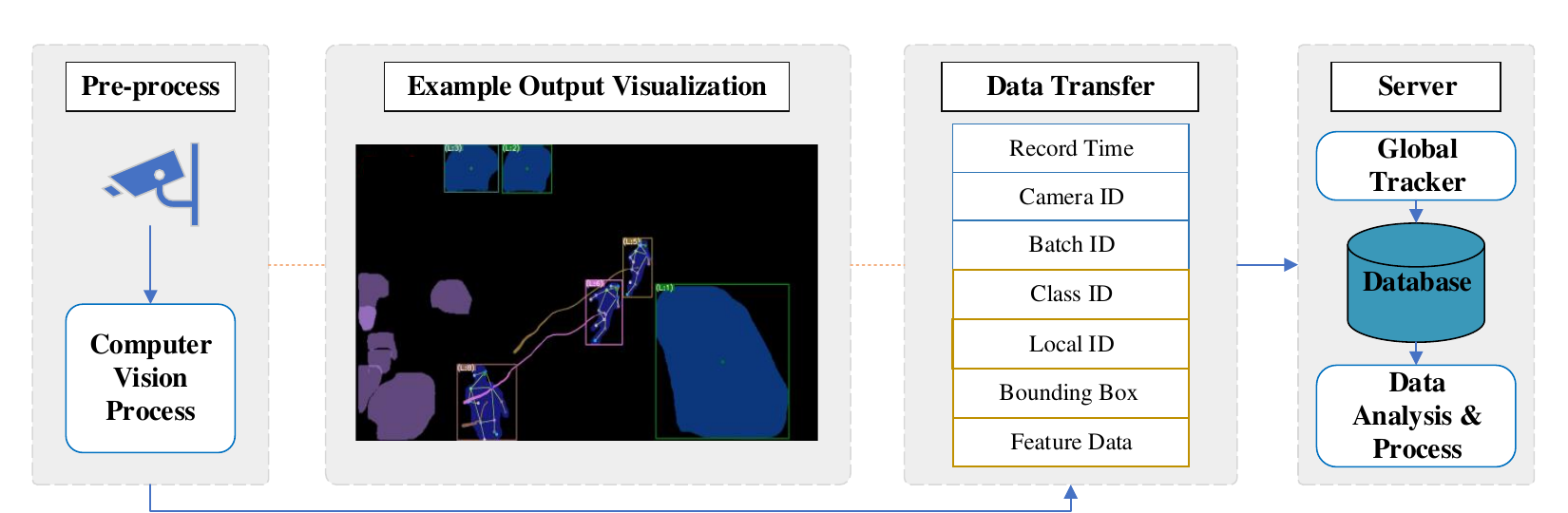}
               \vskip -0.05in
                    \caption{The detailed figure of the system overview. The footage in the Pre-process node is fed to the Computer Vision Pipeline; after the process, the data in the Data Transfer section is extracted and coupled with the Global-IDs will be stored on the database. The data analysis process also takes place on the server.}
                    
                \label{fig:Detailed}
\end{figure*} 
\begin{table*}
\centering
\caption{Visualization of the database at Server}
\label{tab:datastructure}
\begin{tabular}{c|c|c|c|c|c|c}
\rowcolor{DarkGray}
Record Time         & Camera ID & Class ID & Bounding Box  & Feature                  & Local ID & Global ID \\ \hline
2023-10-1 00:00:00  & 1         & 0        & {[}x,y,w,h{]} & Tensors                  & 1        & 1001      \\
2023-10-1 00:00:05  & 2         & 0        & {[}x,y,w,h{]} & Tensors                  & 1        & 1002      \\
...                 & ...       & ...      & ...           & ...                      & ...      & ...       \\
2023-10-20 23:00:00 & 8         & 0        & {[}x,y,w,h{]} & Tensors                  & 5999     & 160001   
\end{tabular}
\end{table*}
\section{Data Structure} \label{Database}

This section delineates the methodology employed for data collection, categorization, and storage in the server database. The process can be broadly divided into three stages: data acquisition, processing via AI pipelines, and storage.

\subsection{Data Acquisition and Processing}
The live stream from each CCTV camera undergoes processing through AI pipelines. Figure \ref{fig:CL} provides an overview of the camera locations used, which consist of 6 indoor cameras and 2 outdoor ones. These cameras were utilized in their existing configuration, and no alterations were made to their specifications. The footage is transferred to the computer vision pipeline using a Wi-Fi connection. The computer vision pipeline utilizes advanced machine learning algorithms to analyze the data \cite{pazho2023ancilia}. As illustrated in Figure \ref{fig:Detailed} these pipelines detect and track various objects, emphasizing humans. Additional data, such as skeletal and feature information, is extracted for the human class. A more granular representation of the data generated from the computer vision methodology before its transfer to the server database is depicted in Figure \ref{fig:Detailed}. This data encompasses the class ID of objects, the local ID of detected entities, bounding box data, and feature data of the human class. Each data packet is tagged with a record time, camera ID, and batch ID. A global tracker utilizes the human feature data to distinguish unique individuals detected across all cameras.

We gathered the data over 8 days, commencing on October 12$^{th}$, 2023, and concluding on October 20$^{th}$, 2023. Our deliberate selection of this timeframe encompassed 4 holidays, including 2 fall break days, 2 weekend days, and 4 working days.

\subsection{Data Storage}
All the data is archived in the database post-global tracking for subsequent analysis and representation. Table \ref{tab:datastructure} elucidates the data storage structure in the server. The columns in the table have the following significance:
\vspace{-5pt}
\begin{itemize}
\item Record Time: Indicates the exact timestamp of data entry. For instance, "2023-10-1 00:00:00" denotes an entry logged at 12:00:00 a.m. on 2023-10-1.

\item Camera ID: Identifies the camera source of the detection.

\item Class ID: A unique identifier for the type of object detected. For example, "0" signifies a human.

\item Bounding Box: Contains coordinates and dimensions of the bounding box around detected objects. It stores the X and Y coordinates of the top left corner of the bounding box as well as the width and height of the bounding box.

\item Feature: Houses the human feature data in Tensors, extracted for global tracking purposes.

\item Local ID: A unique identifier for objects within individual camera streams.

\item Global ID: A universal identifier for objects detected across all cameras, facilitating cross-camera tracking.
\end{itemize}
 As evident from Table \ref{tab:datastructure}, the data it contains does not provide meaningful information to the end user.

\section{Data Representation} \label{Data}

In this section, we explore the detailed data stored in our database. The aim is to transform raw, unprocessed data into insightful and actionable information. Our exposition begins with presenting descriptive data, offering the end user a foundational and general grasp of the dataset's landscape. This preliminary understanding is then augmented with more sophisticated analyses designed to delve deeper into the data's nuances. Through these advanced analytical techniques, we endeavor to magnify the end user's situational awareness, equipping them with a more comprehensive and enriched perspective of the environment.

\subsection{Descriptive Data}\label{Desc.}
In our descriptive data analysis, the linchpin is the Global ID, derived from the human feature data as outlined in Table \ref{tab:datastructure}. This Global ID is a unique identifier, ensuring consistent tracking across various data streams. Five foundational metrics have been presented in this analysis, designed to provide end users with a better understanding of the system's capabilities of the data. Although these data do not provide insightful information to the end user, they can provide a general overview of the traffic flow in the environment over time. 

\textbf{Current Number of People at Real-Time:}
AI pipelines allow the real-time assessment of the number of individuals detected across all cameras. We can determine the current number of people under surveillance by querying the most recent timestamp entries from our database and aggregating the Global IDs. 

\textbf{Hourly Average Number of People Per Camera:}
By grouping the database entries by hour for each camera and averaging the unique Global IDs detected, we can compute the average number of individuals per camera identified hourly. This statistic gives the end user a general understanding of hourly people distribution.  

\textbf{Hourly Average Number of People Per Location:}
Beyond individual cameras, understanding the overall occupation trends in the environment is crucial. By categorizing entries based on the group of cameras and then by hour and subsequently counting and averaging unique Global IDs, we can determine the average traffic across all cameras. This data is incredibly valuable as it offers insights into the anticipated average number of individuals in the environment, leveraging historical data. Such information can prove instrumental for conducting evacuation drills and planning emergency procedures.

\textbf{Total Number of People Over Time:}
The cumulative flow of people over a specified duration provides a broader understanding of traffic flow. By organizing database entries chronologically and counting unique Global IDs, we can chart a time series that depicts the total number of detected individuals, which is a very important metric for controlling the traffic of individuals.

\textbf{Peak Hour Analysis:}
Identifying periods of maximum foot traffic is crucial for various operational decisions. We can pinpoint the peak traffic hours by aggregating data hourly and ranking these hours based on the count of unique Global IDs and use this data to optimize security personnel assignments.

\subsection{Situational Awareness}\label{Aware}

Situational awareness is a key concept in various domains, encapsulating the ability to identify, process, and comprehend critical elements of information about the environment. Situational awareness is about clearly understanding one's surroundings, which is essential for decision-making and proactive responses. Its value enables entities, individuals, or systems to anticipate needs and potential challenges, facilitating timely and informed actions. In the context of surveillance and safety, situational awareness is indispensable. This section delves into four key visualization techniques that support situational awareness: the Occupancy indicator, Statistical anomaly, Bird's eye view, and Heat map. We explain each technique's significance and the underlying methodology employed for its computation. Further, we underscore these visualizations' insights, accentuated with practical examples. When employed judiciously, these techniques serve as powerful tools, shedding light on patterns and anomalies elevating our understanding and proficiency in environment monitoring.

\subsubsection{Occupancy Indicator}\label{OI}
The Occupancy Indicator is a vital tool in video surveillance, aiming to provide contextual understanding concerning the number of people present in a specified location captured by a particular camera. It interprets raw data, such as a head count of seven individuals, into meaningful information by illustrating the relative occupancy level concerning the historical data, whether crowded or within the normal interval. For instance, a count of seven in a small room might signify a high occupancy, whereas in a larger hall, it might denote a low occupancy level. In emergencies like evacuations, knowledge of occupancy levels optimizes response effectiveness. Amid health crises, such as the COVID-19 pandemic, controlling occupancy is essential for public health. By providing a visual or numerical indicator, audiences can grasp the spatial dynamics at a glance.

\begin{algorithm}[tb]
   \caption{Occupancy Indicator Algorithm}
   \label{alg:occupancy}
\begin{algorithmic}
   \State {\bfseries Input:} data frame $df$, historical\_data
   \For{each camera\_id in $df$}
       \For{every 5 seconds in record\_time}
           \State current\_number\_of\_people = length(unique(global\_IDs))
       \EndFor
       \If{current\_number\_of\_people $\leq$ percentile(historical\_data, 25)}
           \State occupancy = "Low Occupancy"
       \ElsIf{current\_number\_of\_people $\leq$ percentile(historical\_data, 75)}
           \State occupancy = "Normal Occupancy"
       \Else
           \State occupancy = "High Occupancy"
       \EndIf
       \State update\_historical\_data(historical\_data, current\_number\_of\_people)
   \EndFor
\end{algorithmic}
\end{algorithm}

Algorithm \ref{alg:occupancy} demonstrates an algorithm to calculate and categorize the occupancy level at different camera locations using the data frame (\texttt{df}), which is continuously updated with new data. The algorithm iterates through each camera ID within the data frame and, for every camera, executes a loop every 5 seconds within a specified range of (\texttt{record\_time}). Each iteration of this nested loop computes the current number of people by determining the length of the set of unique global IDs (\texttt{global\_IDs}) present in the data frame during that time interval.

The occupancy level is then evaluated by comparing the \texttt{current\_number\_of\_people} against the percentiles (25$^{th}$ and 75$^{th}$) from the historical data (\texttt{historical\_data}). Specifically, if the \texttt{current\_number\_of\_people} is less than or equal to the 25$^{th}$ percentile of the historical data, the occupancy is categorized as "Low Occupancy." If it falls between the 25$^{th}$ and 75$^{th}$ percentiles, it is categorized as "Normal Occupancy". Otherwise, if it is greater than the 75$^{th}$ percentile, the occupancy is deemed as "High Occupancy".

Once the occupancy level is determined, the algorithm calls a function \texttt{update\_historical\_data} to update the historical data with the \texttt{current\_number\_of\_people}. This updating step ensures the historical data remains current, allowing for more accurate and relevant occupancy level determinations in subsequent iterations. This process is carried out for every camera ID in the data frame, ensuring a comprehensive evaluation of occupancy levels across all monitored locations. 

\begin{figure}
        \centering
               \includegraphics[width=0.9\linewidth,trim= 0 180 0 0,clip]{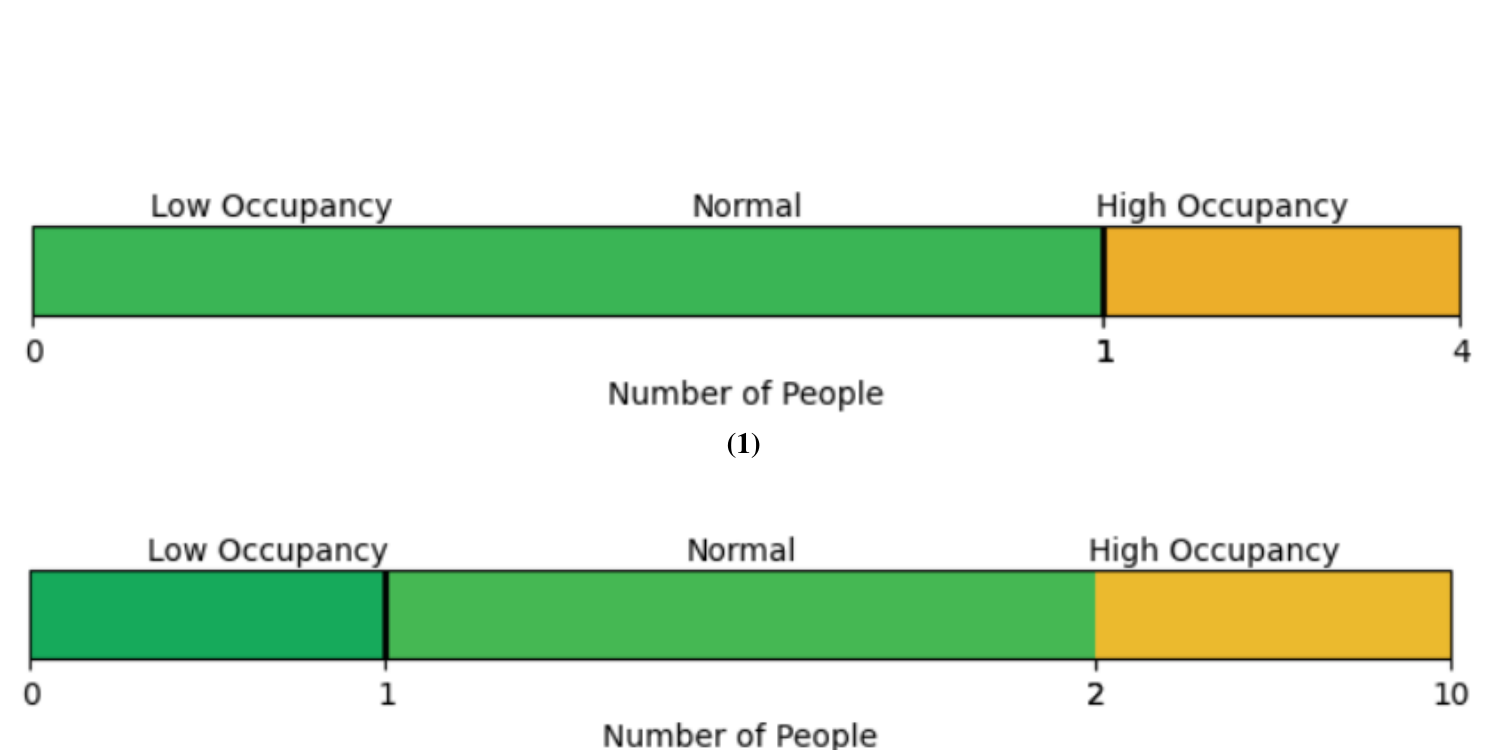}
               \vskip -0.1in
                    \caption{Comparing the Occupancy indicator of the same camera in an hour in two days. The graph (1) shows the Weekend. The graph (2) represents weekday.}
                    
                \label{fig:OI}
\end{figure} 

In Figure \ref{fig:OI}, we can observe a comparison of occupancy indicators for the same camera during the same hour on two distinct days. Graph 1 illustrates the occupancy indicator for a weekend, whereas Graph 2 depicts the occupancy indicator for a weekday. The figure highlights that, during weekends, the detection of 2 people is categorized as "High Occupancy," whereas on weekdays, this number is considered "Normal."

\subsubsection{Statistical Anomaly}\label{SA}

\begin{figure}
        \centering
               \includegraphics[width=1\linewidth,trim= 0 210 0 0,clip]{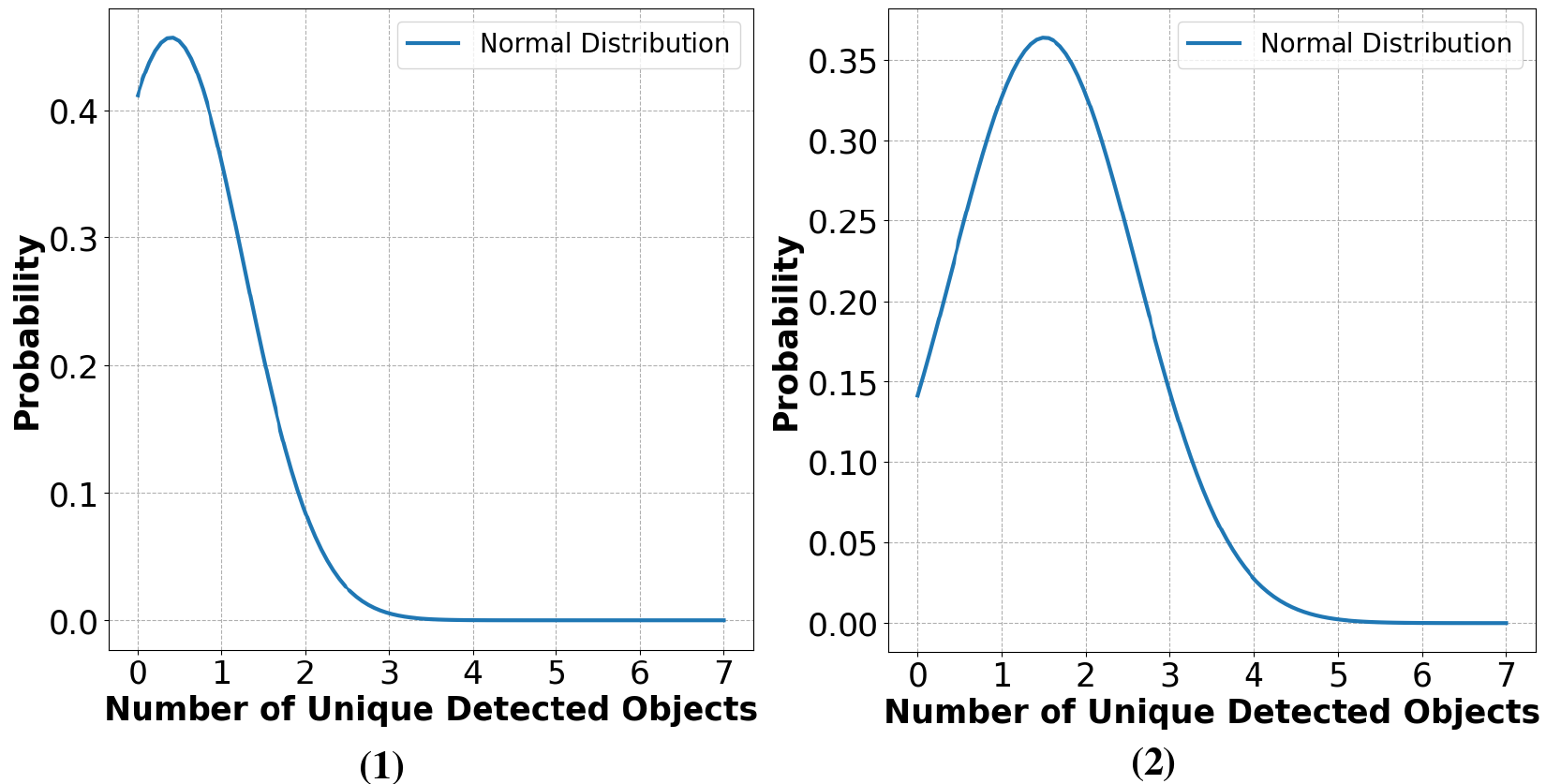}
               \vskip -0.1in
                    \caption{Comparing the normal distributions of the same camera in an hour period in two scenarios. The graph (1) shows the non-detections included in the calculation. The graph (2) represents excluding non-detections.}
                    
                \label{fig:SA_Zero}
\end{figure}

\begin{figure*}
        \centering
               \includegraphics[width=1\linewidth, trim= 0 420 0 0,clip]{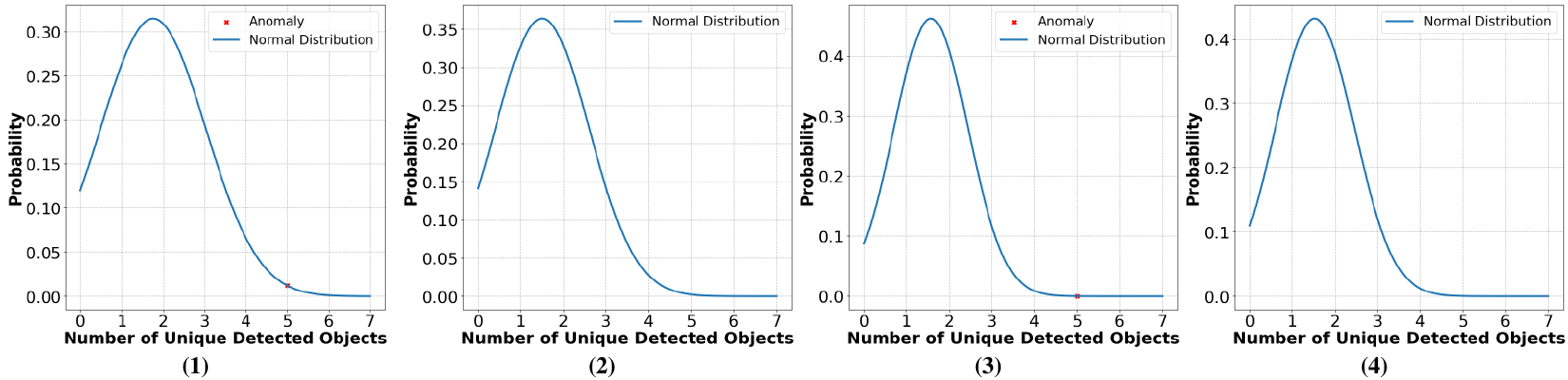}
               \vskip -0.1in
                    \caption{Comparing the normal distributions of the same camera in four scenarios. The graph (1) shows when statistical anomaly happens during weekdays. The graph (2) shows the normal distribution of an hour during weekdays. The graph (3) represents when statistical anomaly happens during weekends. The graph (4) represents the normal distribution of an hour during weekends.}
                    
                \label{fig:SA}
\end{figure*} 

In surveillance and safety, understanding statistical anomalies plays a pivotal role. This is primarily because anomalies, or deviations from the norm, often signal unexpected or unusual events. For instance, a sudden surge in crowd density in monitoring public spaces can indicate potential risks such as unauthorized gatherings, evacuations, or charges. By analyzing historical data, a baseline or 'norm' for crowd density can be established. Any significant deviation from this baseline, especially an increase in crowd density, would be considered an anomaly.

Being able to detect such anomalies in real time enables quick response mechanisms. For safety personnel, it provides an opportunity to proactively address potential threats, ensuring the well-being and security of the public.

\begin{algorithm}[tb]
   \caption{Statistical Anomaly Detection}
   \label{alg:SA}
\begin{algorithmic}
    \State {\bfseries Input:} $df$, $start\_time$, $end\_time$
    \State Filter $df$ for camera ID in the specified time range
    \State Initialize $mean$ to 0 and $std$ to 1
    \State Initialize $detected\_objects$ list
    \For{each 5-second interval from $start\_time$ to $end\_time$}
        \State Count unique detected objects in the interval
        \If{detected objects $>$ 0}
            \State Update $detected\_objects$, $mean$, and $std$
        \EndIf
    \EndFor
\end{algorithmic}
\end{algorithm}

Algorithm \ref{alg:SA} identifies unexpected numbers of detected individuals for each camera relative to historical trends. Recognizing that different locations display varying crowd densities over time, we compute these statistical anomalies hourly. The strategy involves constructing a normal distribution based on historical hourly data. Every 5 seconds, as new detections occur, this distribution is updated. As delineated in Algorithm \ref{alg:SA}, the mean and standard deviation for a specific hour are computed to characterize the data's distribution. Concurrently, current detections are compared against this historical backdrop. If the present number of detections exceeds two standard deviations from the mean, there's a less than 0.05 probability of such an occurrence. This implies that, with 95\% confidence, such an event can be labeled a statistical anomaly.

We've intentionally omitted detections amounting to \texttt{0} in our approach. This decision stems from the rationale that, within a 5-second window, the absence of detections (i.e., detecting no individuals) is more probable than any detection. Including \texttt{0} detections in our computations would skew the results towards \texttt{0}, introducing a bias. This inclusion would result in diminished thresholds for defining anomalies, rendering the statistical analysis less meaningful and effective. Figure \ref{fig:SA_Zero}, compares the normal distributions of the detected objects under these two scenarios for the same camera during the same period. We did not exclude the zeros from our computations in the graph (1). The graph (2) does not include zeros. We can see the skewness of graph (1) toward Zero, resulting in the mean between 0 and 1. In the other scenario, however, the mean is between 1 and 2. As a result, detecting 3 objects in the first scenario is considered an anomaly, while detecting the same number of objects in the second scenario is considered normal.   

In Figure \ref{fig:SA}, we present a comparative analysis of four distinct scenarios captured at two separate temporal intervals of the same camera. Graphs (1) and (2) derive from weekday data. Specifically, Graph (1) illustrates a statistical anomaly, evidenced by detecting 5 individuals within 5 seconds, deviating from the expected probability distribution. In contrast, Graph (2) delineates the typical distribution over an hour-long weekday period. Graphs (3) and (4) are predicated on weekend data, with Graph (3) highlighting another statistical anomaly with 5 people detected. Graph (4) portrays the standard distribution over an hour during weekends. A comparative assessment of Graphs (2) and (4) reveals a higher mean value during weekdays for Graph (2), aligning with anticipated trends.

\subsubsection{Bird's Eye View}\label{BEV}

A bird's eye view, often referred to as a top-down or overhead view, offers a unique vantage point that eliminates perspective distortion commonly associated with ground-level or diagonal images. This perspective allows for accurate spatial representation, ensuring that objects' relative positions and distances from one another are preserved \cite{liu2023bird}. In surveillance or monitoring, a bird's eye view ensures that the entire area of interest is observed without any hidden spots or overlapping regions, a feature often compromised in traditional camera views due to their limited field of view and perspective distortion \cite{dubos2023bird}.

Understanding how people are scattered in a specific area is crucial for various applications such as crowd management, security, and area planning. Accurate representation of people's positions helps determine crowd densities, identify potential choke points, and facilitate effective emergency responses. By utilizing a bird's eye view, these patterns can be more easily discerned, leading to better decision-making and prediction of crowd behaviors.

\begin{algorithm}[tb]
   \caption{Bird's Eye View Transformation}
   \label{alg:birdseye}
\begin{algorithmic}
   \State {\bfseries Input:} data frame $df$, camera\_width, camera\_height, min\_teta, max\_teta
   \State Compute normalized values: $df['normalized\_W']$, $df['normalized\_H']$
   \For{each object in $df$}
       \State Compute $scale\_factor$ using $normalized\_H$
       \State Calculate centroid $C\_X$ and $C\_Y$
       \State Compute $BirdsEye\_X$ and $BirdsEye\_Y$ using $scale\_factor$ and normalized values
   \EndFor
   \For{$camera\_id$ from 1 to 8}
       \State Define start\_time and end\_time
       \State Filter $df$ based on $camera\_ID$ and time range
   \EndFor
\end{algorithmic}
\end{algorithm}
\vskip -0.1in

In Algorithm \ref{alg:birdseye}, a systematic approach is taken to calculate the Bird's Eye View coordinates for objects detected by cameras of the model AXIS P3225-VE Mk II Network Camera \footnote{https://www.axis.com}. The primary objective of this transformation is to project the detected objects onto a top-down view, simulating an overhead perspective.

The first step involves normalizing the object's width and height by dividing them by the camera's resolution parameters, namely, \texttt{camera\_width} and \texttt{camera\_height}. This normalization ensures that the object dimensions are represented as scale-independent relative to the camera's resolution.

Next, a scale factor is computed for each object, a function of its normalized height and the angular field of view parameters, \texttt{min\_teta} and \texttt{max\_teta}. Notably, these angular parameters, \texttt{min\_teta} and \texttt{max\_teta}, are derived directly from the technical specifications provided by the camera producer. The scale factor is crucial for adjusting the object's dimensions in the Bird's Eye View, ensuring that objects farther away appear smaller than those closer to the camera, thereby preserving depth perception.

\begin{figure}
        \centering
               \includegraphics[width=1\linewidth, trim= 0 190 5 0,clip]{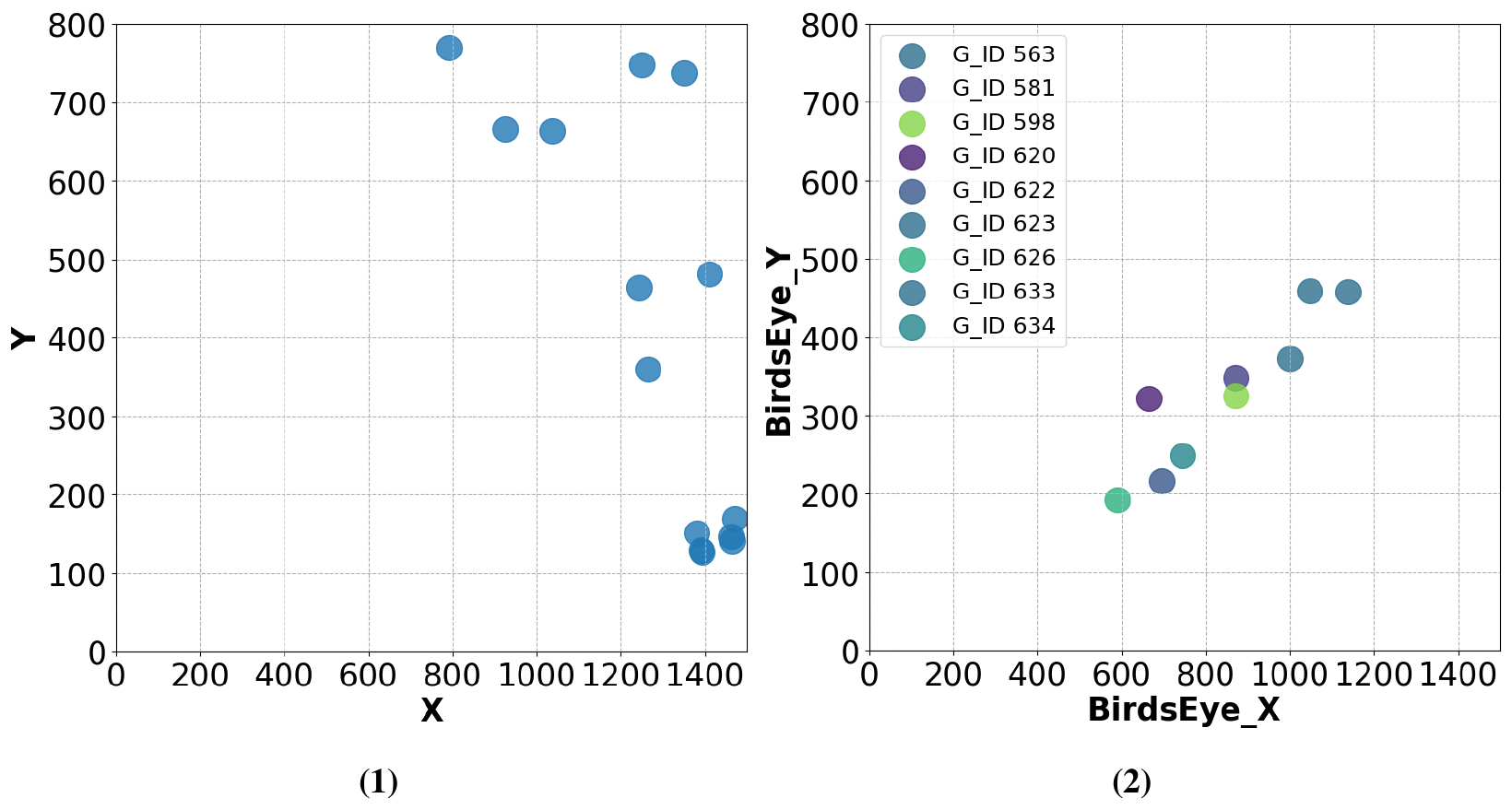}
               \vskip -0.1in
                    \caption{Comparing different views of camera 8 from 2023-10-17 12:17:42 to 12:17:47. The graph (1) generated directly from database shows the original camera view. The graph (2) represents the average birdseye view coordinates.}
                \label{fig:view_graph}
\end{figure} 

The centroid of each object is then determined by calculating the average of its width and height dimensions. This centroid represents the object's central point, essential for accurate positioning in the Bird's Eye View.

Finally, the Bird's Eye View coordinates, \texttt{BirdsEye\_X} and \texttt{BirdsEye\_Y}, are calculated by applying the previously computed scale factor to the object's centroid coordinates. The result of this transformation is an accurate representation of the object's position in a top-down perspective.

The algorithm also includes a data filtering step, where records are filtered based on specific camera IDs and a specified time frame. Therefore, in every 5-second interval, we calculated the unique number of \texttt{global\_ID}s, and based on that, we captured the number of detected objects.  This ensures that only relevant data for a given camera and time range is considered for further analysis.

Figure \ref{fig:view_graph} presents a comparative representation between the original perspective and the processed birdseye view from Camera 8, captured concurrently. The initial graph displays approximately 13 data points within a five-second interval. Post-processing, the bounding boxes corresponding to identical global IDs are averaged, resulting in a birdseye representation of nine distinct individuals. This transformation underscores the efficacy of the bird's eye view process in refining and consolidating data for enhanced clarity and precision.

\subsubsection{Heat map}\label{Heat}

Heatmaps serve as an intuitive visual representation of data distribution across a specified area. In monitoring and surveillance, heatmaps are particularly useful in understanding area congestion and assessing how individuals utilize space over time. A heatmap can instantly reveal high-traffic zones, potential bottlenecks, and less frequented regions by visually representing the frequency of occurrences in different parts of an area. The gradient of colors, usually transitioning from cool to warm tones, indicates the intensity of activity or congestion, with warmer regions signifying higher concentrations of appearance.

\begin{figure*}
        \centering
               \includegraphics[width=1\linewidth, trim= 0 255 0 0,clip]{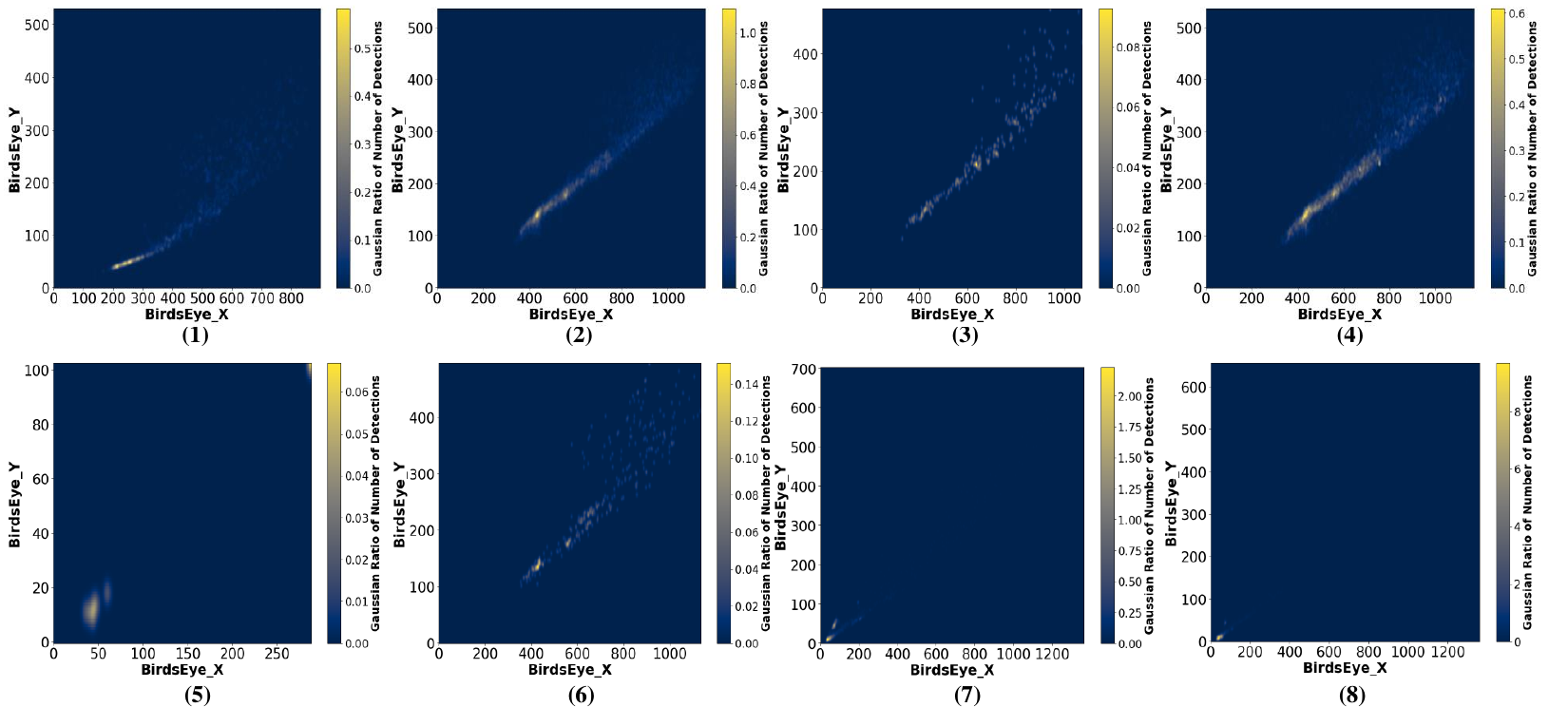}
               \vskip -0.1in
                \caption{Represent four scenarios gradient heatmap of birdseye view in 24 hours. The graph (1), (2), (5), and (6) compare the same day heatmap of four cameras during weekdays. The graph (3) and (4), (7) and (8) compare heatmap of weekdays and weekends of the same camera.}
                    
                \label{fig:Heatmap}
\end{figure*} 
The human brain is trained at processing visual information, and heatmaps capitalize on this inherent capability \cite{hebart2023things}. According to studies in cognitive psychology, visualizations like heatmaps are processed up to 60,000 times faster than text-based data \cite{zeng2023cognition}. Furthermore, using color gradients in heatmaps activates our brain's complicated activating system, making it easier to identify patterns, trends, and anomalies quickly \cite{gireesh2023deep}. The color-coded nature of heatmaps is particularly effective because humans have an innate sensitivity to differences in color, which is evolutionarily rooted in our need to discern varying objects in our environment \cite{najle2023stepwise}. By transforming raw data into a color-coded graphical representation, heatmaps allow for immediate comprehension and pattern recognition. This proximity facilitates the decision-making process and reduces the cognitive load, ensuring stakeholders can make informed decisions without being overwhelmed by raw data.

\begin{algorithm}[tb]
   \caption{Heatmap Generation}
   \label{alg:heatmap}
\begin{algorithmic}
   \State {\bfseries Input:} data frame $df$, start\_date, end\_date
   \State Compute date\_range from start\_date to end\_date
   \For{each date in date\_range}
       \For{camera\_id from 1 to 8}
           \State Filter $df$ for current date and camera\_id
           \If{filtered data is not empty}
               \State Determine max\_x and max\_y from filtered data
               \State Initialize heatmap array of size (max\_y + 1, max\_x + 1)
               \For{each row in filtered data}
                   \State Increment heatmap value at position (Bird's Eye Y, Bird's Eye X)
               \EndFor
           \EndIf
       \EndFor
   \EndFor
\end{algorithmic}
\end{algorithm}
\vskip -0.1in

Algorithm \ref{alg:heatmap} is developed to generate heatmaps based on the data from cameras. We used the calculated Bird's Eye View coordinates to generate heat maps. We processed data for each of the eight cameras, filtering records for every 24-hour window within the specified date range. 

The initial step involves defining a date range between a given \texttt{start\_date} and an \texttt{end\_date}. The algorithm then iterates over this date range and, for each date, processes data from each of the eight cameras. This systematic approach ensures comprehensive coverage, capturing the spatial dynamics of the monitored area over time.
\vskip -0.1in
For each date and camera combination, the dataset, \( df \), is filtered to include only records that fall within the current date and are associated with the current camera. Once filtered, the algorithm checks if there's any data available for processing.

If data is available for the given date and camera, the maximum Bird's Eye View coordinates, \( \texttt{max\_x} \) and \( \texttt{max\_y} \), are determined. These values define the dimensions of a 2D heatmap array, which is initialized with zeros. As the algorithm iterates over each row of the filtered data, the heatmap values are incremented based on the Bird's Eye View positions of detected objects.  Thus, by the end of this process, we obtain a series of heat maps, each capturing the spatial distribution of activity for a specific camera on a particular day, offering invaluable insights into the daily patterns of area usage and congestion. 
We applied Gaussian smoothing to the original heatmaps to enhance the clarity of the heatmaps.

Figure \ref{fig:Heatmap} showcases four distinct gradient heatmaps, each representing a 24-hour interval. Graphs (1), (2), (5), and (6) elucidate pedestrian movement patterns across four separate weekday cameras. The Gaussian ratios reveal varying traffic densities in the four cameras. Furthermore, the distinct patterns illustrate differences in pedestrian behavior when utilizing these areas.  In contrast, the graph (3) and (4), (7) and (8) compares heatmaps from weekdays and weekends. Notably, while the general pattern of the crowd traffic is the same on weekends compared to weekdays, during the weekends, there's a considerable reduction in the Gaussian intensity ratio compared to the heatmap of a typical weekday. For Graph (4) and (8), the Gaussian ratio are 0.6 and 8 during weekdays, but the ratio drops to 0.08 and 2 in weekends shown in Graph (3) and (7). The consistent traffic pattern between the two figures suggests that the facility planning aligns with people's needs. However, the varying Gaussian ratios indicate differences in area occupancy between weekends and weekdays.

\section{Discussions} \label{Discuss}
This paper presents an in-depth analysis of the data representation techniques and visualization methods applied to surveillance data generated by computer vision models on the legacy infrastructure. We summarized the findings of our analysis and provided insights into how these techniques can be leveraged to enhance situational awareness and safety. 
\vskip -0.1in
\subsection{Summary of Findings}

Our analysis began with descriptive data metrics, providing a foundational understanding of the AI surveillance data. We computed key metrics such as the current number of people in real-time, hourly average counts per camera and location, total counts over time, and identified peak traffic hours. These metrics serve as essential building blocks for understanding the dynamics of pedestrian movement in the monitored environment.

We then delved into situational awareness techniques, which play a pivotal role in surveillance and safety. The Occupancy Indicator, utilizing real-time data, categorizes occupancy levels as low, normal, or high. This is crucial for ensuring adherence to safety regulations, emergency response, and managing public spaces during health crises like the COVID-19 pandemic.

The Statistical Anomaly detection algorithm identifies deviations from historical crowd density patterns, enabling real-time detection of unexpected events. This is invaluable for proactive responses to potential threats, ensuring public safety, and crowd management.

The Bird's Eye View transformation provides a top-down perspective of the monitored area, allowing for accurate spatial representation of individuals. This view is essential for crowd management, security, and facility planning, as it helps identify congestion, choke points, and crowd behavior patterns.

Heatmaps visually represent data distribution over time, highlighting high-traffic areas and congestion patterns. This is particularly useful for understanding area utilization, identifying bottlenecks, optimizing resource allocation, enhancing safety measures, and improving overall spatial planning in various environments.

\subsection{Insights for Real-World Applications}

The application of these techniques to AI surveillance data has several implications for safety, health, transportation, architecture and urban planning:

Emergency Response: The Occupancy Indicator and Statistical Anomaly Detection are vital for emergency response scenarios. They can trigger alerts in case of overcrowding or unexpected events, ensuring a rapid and coordinated response to emergencies.

Public Health: Maintaining optimal occupancy levels is critical during health crises, such as pandemics. The Occupancy Indicator, along with real-time monitoring, helps enforce social distancing measures, contributing to public health and safety.

Crowd Management: Bird's Eye View and Heatmaps aid in crowd management by identifying high-traffic areas and congestion patterns. Urban planners can use this information to optimize the layout of public spaces, transportation systems, and event venues.

Resource Allocation: Real-time data analysis enables efficient resource allocation. For example, law enforcement agencies can deploy personnel to areas with high occupancy or potential anomalies, improving resource utilization.

Predictive Analysis: Historical data and anomaly detection can be used for predictive analysis. By identifying patterns in crowd behavior and anomalies, authorities can anticipate and mitigate potential safety risks.

Urban Design: Insights from Bird's Eye View and Heatmaps can influence urban design decisions. Planners can create more pedestrian-friendly environments, optimize traffic flow, and design public spaces that cater to the community's needs.

Data-Driven Decision-Making: These techniques empower decision-makers with data-driven insights. Whether optimizing public transportation routes or managing event crowd flow, data-based decisions lead to better outcomes.

\subsection{Future Work}

Several promising avenues for future research and development emerge in the realm of safety and urban planning utilizing AI surveillance data. First, integrating AI surveillance data with IoT sensors offers the potential for comprehensive environmental monitoring, encompassing aspects like air quality, temperature, and noise levels, thereby enriching safety and planning efforts. Second, dynamic resource allocation algorithms could automatically dispatch emergency responders or adjust real-time traffic signals based on surveillance data, enhancing responsiveness in critical situations. Moreover, analyzing human behavior in crowded areas using computer vision techniques can help identify suspicious activities or signs of distress, improving safety. Additionally, creating simulation models for urban environments allows testing of various planning scenarios, while privacy-preserving techniques ensure responsible data usage. Lastly, integrating multi-modal data sources, including transportation and weather data, provides a holistic view of urban dynamics for more informed decision-making.

\section*{Acknowledgment}
This research is supported by the National Science Foundation (NSF) under Award No. 1831795.

\bibliographystyle{unsrtnat}
\bibliography{references}  






\end{document}